\documentclass[article]{elsarticle}
\usepackage{amsmath, amsthm, amssymb, amsfonts}
\usepackage{ragged2e}
\journal{Physica scripta}
\begin{document}
\begin{frontmatter}
\title{Rogue waves in space dusty plasmas}
\author{N. A. Chowdhury$^*$ and A. A. Mamun}
\address{Department of Physics\\
Jahangirnagar University\\
Savar, Dhaka-1342, Bangladesh\\
Email:$^*$nurealam1743phy@gmail.com}
%%%%%%%%%%%%%%%%%%%%%%%%%%%%%%%%%%%%%%%%%%%%%%%%%%%%%%%%%%%%%%%%%%%%%%%%%%%%%%%%%%%%%%%%%%%%%%%%%%%%%%%%%%%%%%%%%%%%%%%%%%%%%%%%%%%%%
\begin{abstract}
  The modulational instability of dust-acoustic (DA)  waves (DAWs), and corresponding DA rogue waves (DARWs) in a realistic space dusty plasma system (containing inertial warm  positively and
  negatively charged dust, isothermal ions, and nonthermal kappa distributed electrons) has been theoretically investigated. The nonlinear Schr\"{o}dinger (NLS) equation is derived by using reductive perturbation method for this investigation. It is observed that the  dusty plasma system under consideration supports two modes, namely fast and slow modes, and that both of these
  two modes can be stable or unstable depending the sign of ratio of the dispersive and nonlinear coefficients. The numerical analysis have shown that the basic
  features (viz. stability/instabilit, growth rate, amplitude, etc.) of the DAWs associated with the fast mode are significantly modified by superthermal parameter ($\kappa$) and  other various plasma parameters.
  The results of our present investigation should be useful for understanding DARWs in space plasma systems, viz. mesosphere and ionospehre.
\end{abstract}
\end{frontmatter}
%%%%%%%%%%%%%%%%%%%%%%%%%%%%%%%%%%%%%%%%%%%%%%%%%%%%%%%%%%%%%%%%%%%%%%%%%%%%%%%%%%%%%%%%%%%%%%%%%%%%%%%%%%%%%%%%%%%%%%%%%%%%%%%%%%%%%%%%
\section{Introduction}
\normalsize
Empirical results disclosed the existence of charged dust in space (viz.
in lower and upper mesosphere, cometary tails, planetary rings, planetary magnetospheres, inter-planetary
spaces, interstellar media, etc.\cite{Shukla2002,Verheest2000,Shukla2001,Mendis1994}) and laboratory plasmas. Nowadays physicists are exclusively using wave dynamics
like, dust-ion acoustic waves (DIAWs), DAWs and DARWs in understanding electrostatic
density perturbations and potential structures (viz. soliton, shock, vortices, and rogue profile, etc.)
that are observed in dusty plasma  system \cite{Gill2010,Tasnim2013}. DAWs which arises due to the inertia of the the dust mass
and restoring force is provided by the thermal pressure of ions and electrons. Many researchers
have also observed both theoretically and experimentally that the existence of gigantic and highly
charged dust grains in plasma can significantly modify the behaviour of the usual waves and stabilities.

A set of authors observed the electrostatic linear/nonlinear structures in dusty plasma system by
considering only negatively charged dust, ions, and electrons in their considered
plasma model. The presence of positively charged dust along with negatively charged dust  has also been observed
in different regions of space; viz. upper mesosphere, cometary tails,  Jupiter’s magneto-sphere,
etc \cite{Mendis1994,Chow1993,Chow1994,Mamun2002}. There are three \cite{Fortov1998} basic mechanisms (I. photoemission in the presence of a flux of ultraviolet
photons, II. thermionic emission induced by radiative heating, and III. secondary emission of electrons from the surface of the
dust grains.) by which a dust grain becomes positively charged and coexist with negatively charged dust grains plasmas.

In case of natural space or in laboratories plasmas high energy
particles may accompany \cite{Rehman2016} with isothermal distrubuted particles and such kind of high
energetic particles characteristics are deviated from the  renowned Maxwellian distribution due to the external
forces acting on the natural space plasmas or due to wave particle interaction. Sometimes
this type of particles can be governed by non-Maxwellian high-energy tail distribution which is known
as generalized Lorentzian (kappa) distribution \cite{Vasyliunas1968,Summers1991,Mace1995,Maksimovic1997,Vi˜nas2005,Hellberg2009,Baluku2008,Baluku2010,Choi2011,Sultana2011,Shahmansouri2013a}.
The relationship between kappa distribution to the Maxwellian distribution was first introduced by
Vasylinuas \cite{Vasyliunas1968}. In kappa distribution, the parameter $\kappa$ is the measurement of the slope of the energy
spectrum of the superthermal particles forming the tail of the
velocity distribution function which is also called spectral
index. Lower values of $\kappa$ represents a hard spectrum with a
strong non-Maxwellian tail \cite{Summers1991}. The Lorentzian or
kappa distribution function  reduces to the Maxwellian
(thermal equilibrium distribution) for the limit of large
spectral index , i.e., $\kappa
\rightarrow\infty$.

Recently, Sayed and Mamun \cite{Sayed2007} examined  the basic properties of small but finite amplitude
solitary potential in a dusty plasma system consisting of electros, ions and negative as well as positive
dust. They observed that the presence of additional positive dust component has
significantly modified the the basic properties of solitary potential. El-Taibany \cite{El-Taibany2013} Studied the nature
of nonlinear DA solitary waves in four component inhomogeneous
dusty plasma with opposite charge dust grains. Where he found that his considered model revealed two DA velocities,
one slow (compressive and rarefactive are allowed) and other is fast (compressive soliton is created).
The investigations of the MI of waves, both theoretically and experimentally have been exponentially
increasing day by day due to their rigorous successful applications in space, laboratory plasmas,
ocean wave, optics. Recently, a bounce of authors has been used NLS equation to understand different
nonlinear phenomena (MI and rogue waves), which are observed in laboratory and  space plasmas.
Abdelwahed \textit{et al.} \cite{Abdelwahed2016} examined rogue waves characteristics are totally depended on
the ionic density ratio, ionic mass ratio, and superthermal parameter in ion pair
superthermal plasma. Gill \textit{et al.} \cite{Gill2010} studied MI of dust acoustic solitons
in multicomponent plasma with kappa distributed electron and ions, where they observed presence of positive
dust component modify the domain of the MI and localized envelope excitations. The aim of the present paper is, by
using reductive perturbation method  a NLS equation is derived for nonlinear electrostatic  DARWS in unmagnetized
dusty plasmas in the presence of inertial opposite polarity (positively and massive negatively charged) dust,
isothermal ions, and superthermally kappa distributed electrons.

The manuscript is organized as follows: The basic
governing equations of our considered plasma model is presented in sec. \ref{Governing Equations}. By using reductive perturbation technique,
we derive a NLS equation which governs the slow amplitude evolution
in space and time is given in sec. \ref{NLS}.  Modulational instability  and rouge waves are presented, in sec. \ref{Modulational}.
The discussion is  provided in sec. \ref{Discussion}. Finally conclusion is presented in sec.  \ref{Conclusion}.
%%%%%%%%%%%%%%%%%%%%%%%%%%%%%%%%%%%%%%%%%%%%%%%%%%%%%%%%%%%%%%%%%%%%%%%%%%%%%%%%%%%%%%%%%%%%%%%%%%%%%%%%%%%%%%%%%%%%%%%%%%%%%%%%%%%
\section{Governing Equations}\label{Governing Equations}
We consider a collisionless, fully ionized unmagnetized plasma system comprising of
inertial warm negatively  charged massive dust (mass $m_1$, charge $q_1=-Z_1e$) and positively
charged dust (mass $m_2$, charge $q_2=+Z_2e$), as well as kappa distributed electrons
(mass $m_e$; charge $-e$), and isothermal ions (mass $m_i$; charge $+e$). $Z_1~(Z_2)$ is the number of elctrons (protons)
residing on a negative (positive) dust. At equilibrium, the quasi-neutrality condition can be expressed as $n_{10} + n_{e0}=  n_{20}+n_{i0}$,
where $n_{10}$, $n_{e0}$, $n_{20}$ and $n_{i0}$ are the equilibrium number densities of warm negatively
charged massive dust, kappa distributed electron, positively charged dust, and isothermal ion,
respectively. The normalized governing equations of the DAWs in our considered plasma system are given below
\begin{eqnarray}
&&\frac{\partial n_1}{\partial t}+\frac{\partial}{\partial x}(n_1 u_1)=0,\label{eq1}\\
&&\frac{\partial u_1}{\partial t} + u_1\frac{\partial u_1 }{\partial x} + 3\sigma_1  n_1\frac{\partial n_1 }{\partial x}=\frac{\partial \phi}{\partial x},\label{eq2}\\
&&\frac{\partial n_2}{\partial t}+\frac{\partial}{\partial x}(n_2 u_2)=0,\label{eq3}\\
&&\frac{\partial u_2}{\partial t} + u_2\frac{\partial u_2 }{\partial x}+ 3\sigma_2  n_2\frac{\partial n_2 }{\partial x}=-\alpha \frac{\partial \phi}{\partial x},\label{eq4}\\
&&\frac{\partial^2 \phi}{\partial x^2}=(\mu_i+\beta-1)n_e-\mu_i n_i+n_1-\beta n_2. \label{eq5}
\end{eqnarray}
For inertialess kappa distributed electron, we can obtain the expressions for electron  number densities as
\begin{eqnarray}
&&n_e= \left [1-\frac{\delta \phi}{(\kappa -3/2)}\right]^{-\kappa+{1/2}}=1+ \eta_1 \delta \phi+ \eta_2\delta^2  \phi^2 +\eta_3 \delta^3 \phi^3 +\cdot\cdot\cdot \label{eq6}\
\end{eqnarray}
where
\begin{eqnarray}
&&\eta_1=\frac{(\kappa-\frac{1}{2})}{(\kappa-\frac{3}{2})},~~\eta_2=\frac{(\kappa-\frac{1}{2})(\kappa+\frac{1}{2})}{2(\kappa-\frac{3}{2})^2},~~\eta_3=\frac{(\kappa-\frac{1}{2})(\kappa+\frac{1}{2})(\kappa+\frac{3}{2})}{6(\kappa-\frac{3}{2})^3}.\nonumber\
\end{eqnarray}
For inertialess isothermal ion, we can obtain the expressions for ion number densities as
\begin{eqnarray}
&& n_i= \mbox{exp}~(-\phi) =1 -\phi+\frac{\phi^2}{2}-\frac{\phi^3}{6}+\cdot\cdot\cdot\label{eq7}
\end{eqnarray}
substituting Eqs. (\ref{eq6}) and (\ref{eq7}),  into Eq. (\ref{eq5}), and expanding up to third order, we get
\begin{eqnarray}
&&\frac{\partial^2 \phi}{\partial x^2}=(\beta-1)+n_1-\beta n_2+\gamma_1 \phi+\gamma_2\phi^2+\gamma_3 \phi^3\cdot\cdot\cdot \label{eq8}
\end{eqnarray}
Here
\begin{eqnarray}
&&\gamma_1=\eta_1\delta(\mu_i+\beta-1)+\mu_i,~~\gamma_2=\eta_2\delta^2(\mu_i+\beta-1)-\mu_i/2, \nonumber\\
&&\gamma_3=\eta_3\delta^3(\mu_i+\beta-1)+\mu_i/6,\nonumber\
\end{eqnarray}
and

~~~~~~~~~~~~~$\sigma_1=\frac{T_1}{Z_1 T_i}$,~~~~$\sigma_2=\frac{T_2 m_1}{Z_1T_i m_2}$,~~~~~$\alpha=Z\mu$,~~~~~$Z=\frac{Z_2}{Z_1}$,~~~~$\beta=ZR$,

~~~~~~~~~~~~~$\mu=\frac{m_1}{m_2}$,~~~~~~~~$\mu_i=\frac{n_{i0}}{Z_1n_{10}}$,~~~~~~~~$\delta=\frac{T_i}{T_e}$,~~~~~$R=\frac{n_{20}}{n_{10}}$.

\noindent In the above equations, $n_1$ ($n_2$) is the number density of negatively charged massive dust
(positively charged dust) normalized by its equilibrium value
$n_{10}$ ($n_{20}$); $u_1(u_2)$ is the negative (positive) charged dust fluid speed
normalized by $C_1=(Z_1 T_i/m_1)^{1/2}$, and the electrostatic wave potential $\phi$
is normalized by $T_i/e$ (with $e$ being the magnitude of an electron charge).
$T_1$, $T_2$, $T_i$ and $T_e$ is the temperature of  negatively charged massive dust,
positively charged dust, isothermal ion, and kappa distributed  electrons, respectively. The time and space variables are normalized by
${\omega^{-1}_{pd1}}=(m_1/4\pi Z^2_1e^2 n_{10})^{1/2}$ and $\lambda_{Dd1}=(T_{i}/4 \pi Z_1 e^2 n_{10})^{1/2}$, respectively.
%%%%%%%%%%%%%%%%%%%%%%%%%%%%%%%%%%%%%%%%%%%%%%%%%%%%%%%%%%%%%%%%%%%%%%%%%%%%%%%%%%%%%%%%%%%%%%%%%%%%%%%%%%%%%%%%%%%%%%%%%%%%%%%%%%%%%%%%%%%%%%%%%%%%%%%%%%%%%
\begin{figure*}[htp]
  \centering
  \begin{tabular}{cccc}
    \includegraphics[width=130mm]{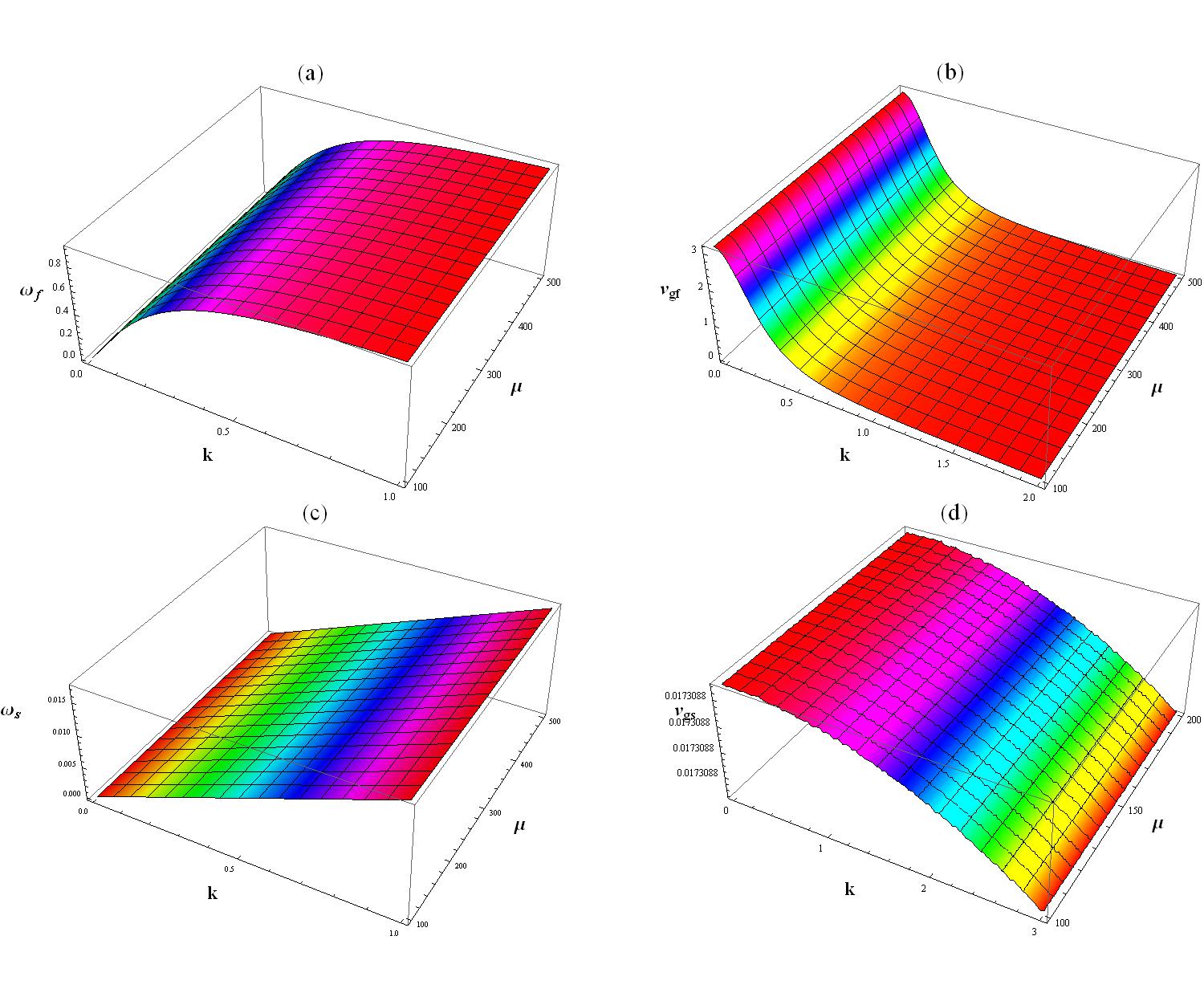}&
  \end{tabular}
  \label{figur}\caption{Showing the variation of $\omega$  against $k$ and $\mu$,
  (a) when $\omega$ is fast mode, and (c) when $\omega$ is slow mode.
  Showing the variation of $v_g$  against $k$ and $\mu$, (b) when $\omega$ is fast mode, and
  (d) when $\omega$ is slow mode. Generally, all the figures are generated by using these values $Z=0.01$, $\mu=150$, $R=0.1$,
  $\mu_i=0.4$,  $\delta=0.3$,  $\kappa=3$, $\sigma_1=0.00001$, and $\sigma_2=0.0001$.}
\end{figure*}

\begin{figure*}[htp]
  \centering
  \begin{tabular}{cccc}
  \includegraphics[width=60mm]{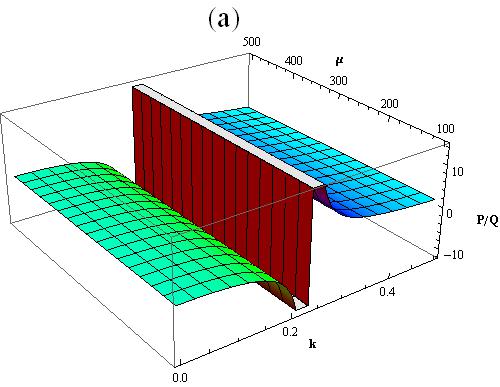}&
   % \hspace{0.25in}

  \includegraphics[width=60mm]{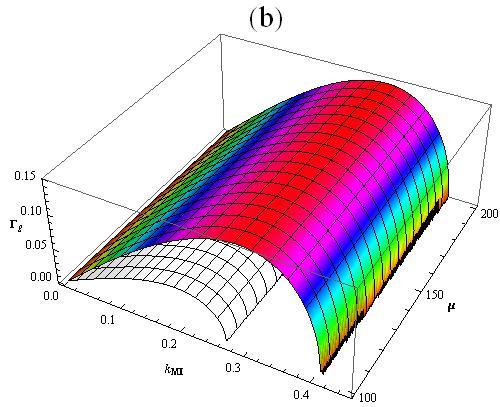}\\
   % \hspace{0.25in}

  \includegraphics[width=60mm]{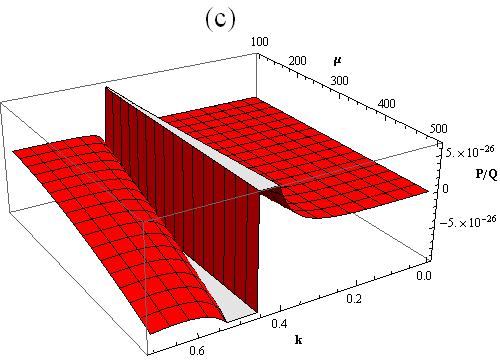}&
   %\hspace{0.25in}

  \includegraphics[width=60mm]{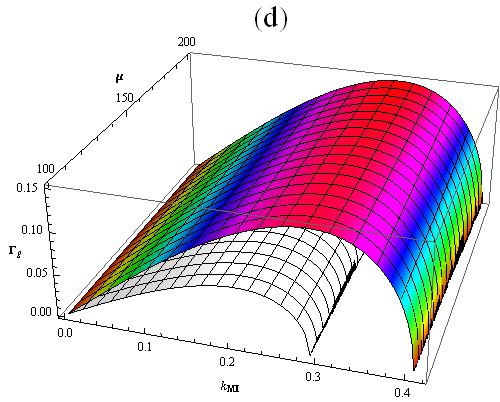}\\
  \end{tabular}
  \label{figur}\caption{Showing the variation of $P/Q$ against $k$ and $\mu$, (a) when $\omega$ is fast mode, and (c) when $\omega$ is slow mode.
   Plot of the of MI growth rate $(\Gamma_g)$ against $K_{MI}$ and $\mu$, and (b) for $\kappa=3$ (Hue curve), $\kappa=30$ (Gray curve),
  (d) for  $\delta=0.2$ (Hue curve), $\delta=0.3$ (Gray curve), respectively. Along with $\omega$ is fast mode, $k=0.4$, and $\Phi_0=0.5$.}
\end{figure*}

\begin{figure*}[htp]
  \centering
  \begin{tabular}{cccc}
    \includegraphics[width=130mm]{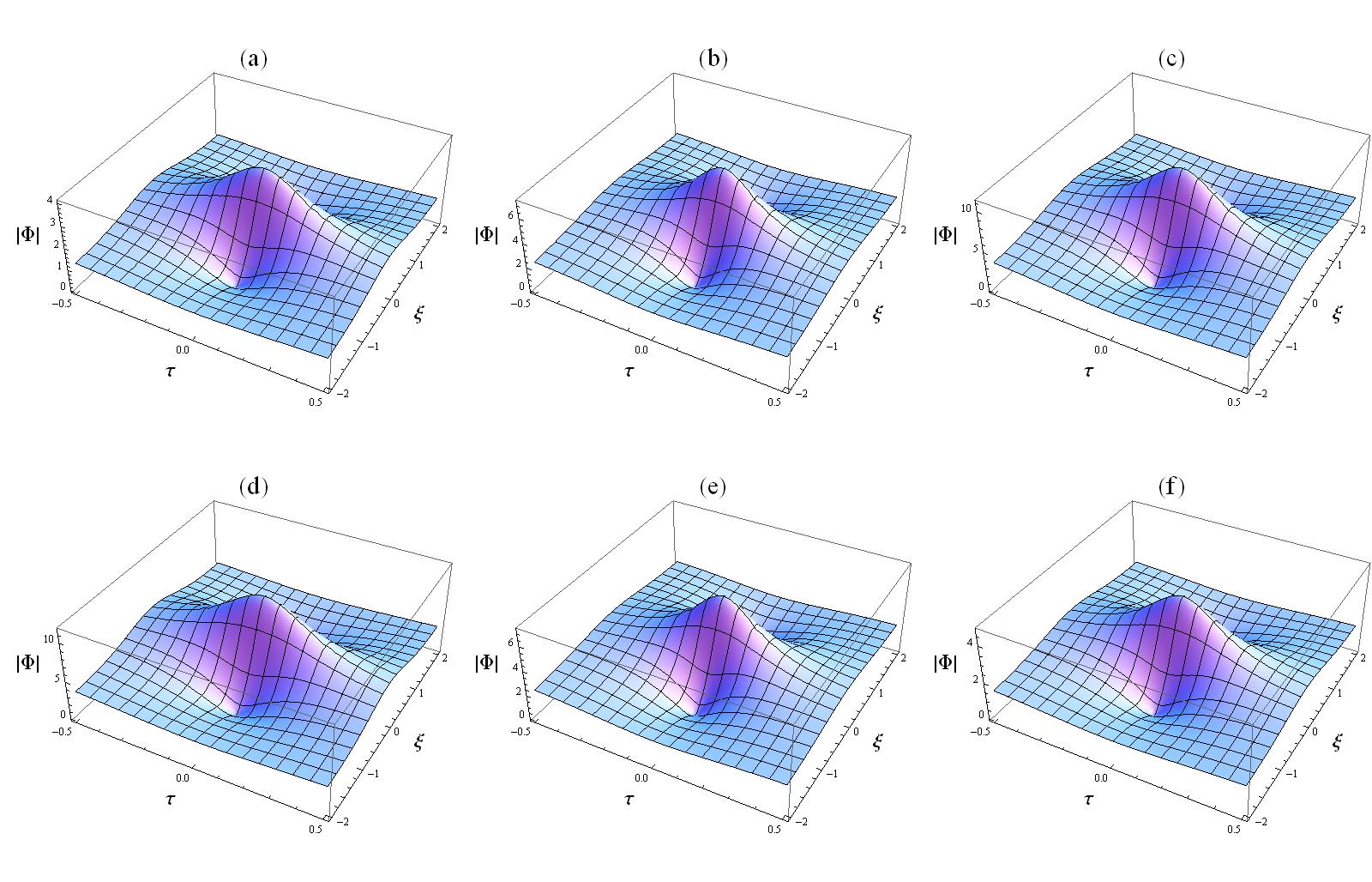}&
  \end{tabular}
  \label{figur}\caption{Showing the variation of $|\Phi|$ against $\xi$ and $\tau$,
  (a) For $\kappa=2.5$, (b) For $\kappa=3$, (c) For $\kappa=30$, (d) For $\delta=0.1$,
  (e) For $\delta=0.3$, and (f) For $\delta=0.35$. Along with $\Phi_0=0.5$, $k=0.4$ and $\omega$ is fast mode.}
\end{figure*}
\begin{figure*}[htp]
  \centering
  \begin{tabular}{cccc}
    \includegraphics[width=130mm]{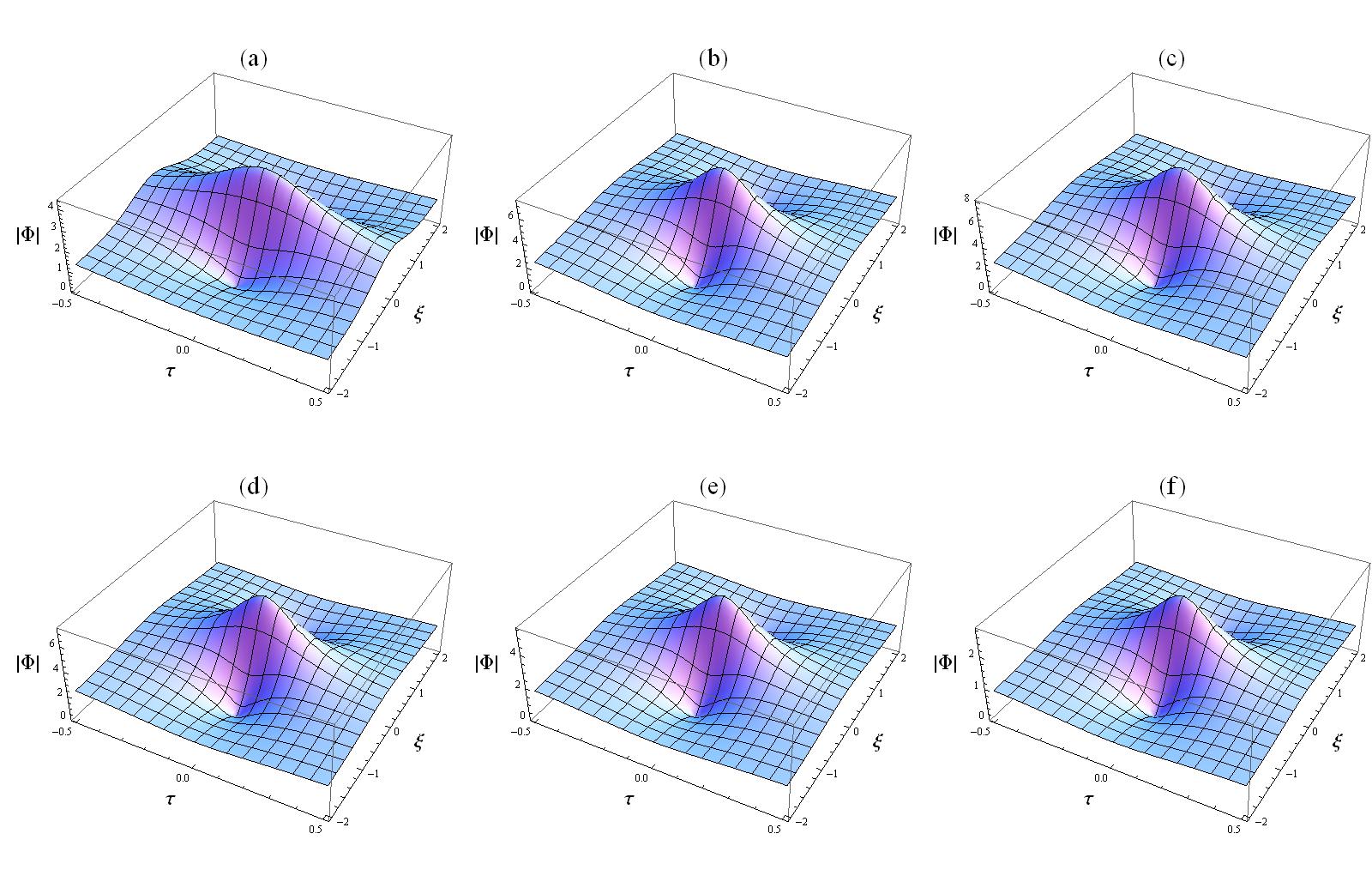}&
  \end{tabular}
  \label{figur}\caption{Showing the variation of $|\Phi|$ against $\xi$ and $\tau$,
  (a) For $\mu_i=0.35$, (b) For $\mu_i=0.40$, (c) For $\mu_i=0.45$, (d) For $Z=0.01$,
  (e) For $Z=0.04$, and (f) For $Z=0.07$. Along with $\Phi_0=0.5$, $k=0.4$ and $\omega$ is fast mode.}
\end{figure*}
%%%%%%%%%%%%%%%%%%%%%%%%%%%%%%%%%%%%%%%%%%%%%%%%%%%%%%%%%%%%%%%%%%%%%%%%%%%%%%%%%%%%%%%%%%%%%%%%%%%%%%%%%%%%%%%%%%%%%%%
\section{Derivation of the NLS Equation}\label{NLS}
To study the MI of the DAWs, we will derive the NLS equation by
employing the reductive perturbation method. So we first introduce
the independent variables are stretched as
\begin{eqnarray}
&&\xi={\epsilon}(x  - v_gt),~~~~~~~~\tau={\epsilon}^2 t, \label{eq9}
\end{eqnarray}
where $v_g$ is the envelope group velocity to be determined later and $\epsilon ~(0<\epsilon<1)$ is a
small (real) parameter. Then we can
write a general expression for the dependent variables as
\begin{eqnarray}
&&n_1=n_{10} +\sum_{m=1}^{\infty}\epsilon^{(m)}\sum_{l=-\infty}^{\infty}n_{1l}^{(m)}(\xi,\tau)~\mbox{exp}[i l(kx-\omega t)], \nonumber\\
&&u_1=\sum_{m=1}^{\infty}\epsilon^{(m)}\sum_{l=-\infty}^{\infty}u_{1l}^{(m)}(\xi,\tau)~\mbox{exp}[i l(kx-\omega t))], \nonumber\\
&&n_2=n_{20} +\sum_{m=1}^{\infty}\epsilon^{(m)}\sum_{l=-\infty}^{\infty}n_{2l}^{(m)}(\xi,\tau)~\mbox{exp}[i l(kx-\omega t)], \nonumber\\
&&u_2=\sum_{m=1}^{\infty}\epsilon^{(m)}\sum_{l=-\infty}^{\infty}u_{2l}^{(m)}(\xi,\tau)~\mbox{exp}[i l(kx-\omega t)], \nonumber\\
&&\phi=\sum_{m=1}^{\infty}\epsilon^{(m)}\sum_{l=-\infty}^{\infty}\phi_{l}^{(m)}(\xi,\tau)~\mbox{exp}[i l(kx-\omega t)], \label{eq10}
\end{eqnarray}
where k and $\omega$ are real variables representing the fundamental (carrier) wave number
and frequency, respectively.  $ M_l^{(m)}$ satisfies the pragmatic condition $ M_l^{(m)}= M_{-l}^{(m)^*}$, where the asterisk denotes
the complex conjugate. The derivative operators in the above equations are treated as follows:
\begin{eqnarray}
&&\frac{\partial}{\partial t}\rightarrow\frac{\partial}{\partial t}-\epsilon v_g \frac{\partial}{\partial\xi}+\epsilon^2\frac{\partial}{\partial\tau},~~~~~~~~~\frac{\partial}{\partial x}\rightarrow\frac{\partial}{\partial x}+\epsilon\frac{\partial}{\partial\xi}. \label{eq11}
\end{eqnarray}
Substituting Eqs. (\ref{eq10}) and (\ref{eq11}) into equations Eqs. $(\ref{eq1})-(\ref{eq4})$ and (\ref{eq8}) and collecting power terms of $\epsilon$,
first-order  $(m=1)$ equations with  $l=1$
\begin{eqnarray}
&&-i\omega n_{11}^{(1)}+iku_{11}^{(1)}=0,~~~~~~~ik\lambda n_{11}^{(1)}-i\omega u_{11}^{(1)}-ik \phi_1^{(1)}=0,\nonumber\\
&&-i\omega n_{21}^{(1)}+iku_{21}^{(1)}=0,~~~~~~~ik\theta n_{21}^{(1)}-i\omega u_{21}^{(1)}+ik \alpha\phi_1^{(1)}=0,\nonumber\\
&&-n_{11}^{(1)}-k^2\phi_1^{(1)}-\gamma_1\phi_1^{(1)}+\beta n_{21}^{(1)}=0,\label{eq12}
\end{eqnarray}
where $\lambda=3 \sigma_1$ and $\theta=3 \sigma_2$. The solution for the first harmonics read as
\begin{eqnarray}
&& n_{11}^{(1)}=\frac{k^2}{S}\phi_1^{(1)},~~~~~~~~u_{11}^{(1)}=\frac{k\omega}{S}\phi_1^{(1)},\nonumber\\
&&n_{21}^{(1)}=\frac{\alpha k^2}{A}\phi_1^{(1)},~~~~~~u_{21}^{(1)}=\frac{\alpha k\omega}{A}\phi_1^{(1)},\label{eq13}
\end{eqnarray}
where $S=\lambda k^2-\omega^2$ and $A=\omega^2-\theta k^2$. We thus obtain the dispersion relation for DAWs
\begin{eqnarray}
&&\omega^2=\frac{k^2M\pm k^2 \sqrt{M^2-4 GH}}{2G},\label{eq14}
\end{eqnarray}
where, $M=(\theta k^2+\lambda k^2+\theta \gamma_1+\lambda\gamma_1+\alpha\beta+1), G=(k^2+\gamma_1 )$, and
$H=(\theta\lambda k^2+\theta\gamma_1\lambda+\theta+\alpha \beta\lambda).$
In Eq. (\ref{eq14}), to get real and positive value of $\omega$, the condition $M^2>4GH$ should be verified.
The positive (negative) sign corresponds to the fast (slow)  modes, respectively.
The second-order when $(m=2)$ reduced equations with $(l=1)$ are
\begin{eqnarray}
&&n_{11}^{(2)}=\frac{k^2}{S}\phi_1^{(2)}+\frac{i}{S^2}(\lambda k^3-2v_g\omega k^2+k\omega^2-kS )\frac{\partial \phi_1^{(1)}}{\partial\xi},\nonumber\\
&&u_{11}^{(2)}=\frac{k\omega}{S}\phi_1^{(2)}+\frac{i}{S^2}(\lambda\omega k^2-2v_g k\omega^2+\omega^3-v_g k S)\frac{\partial\phi_1^{(1)}}{\partial\xi},\nonumber\\
&&n_{21}^{(2)}=\frac{\alpha k^2}{A}\phi_1^{(2)}-\frac{i\alpha}{A^2}(\theta k^3-2\omega v_g k^2+k\omega^2+k A)\frac{\partial\phi_1^{(1)}}{\partial\xi},\nonumber\\
&&u_{21}^{(2)}=\frac{\alpha k\omega}{A}\phi_1^{(2)}-\frac{i\alpha}{A^2}(\theta\omega k^2-2v_g k\omega^2+\omega^3+v_g k A)\frac{\partial\phi_1^{(1)}}{\partial\xi}, \nonumber\
\end{eqnarray}
with the compatibility condition
\begin{eqnarray}
&&v_g=\frac{k^2(\lambda A^2 + \alpha \beta \theta S^2)+ \omega^2 (A^2+\alpha \beta S^2)-2 S^2 A^2-SA(A-\alpha \beta S )}{2k \omega( A^2 + \alpha \beta S^2 )},\nonumber\
\end{eqnarray}
The second-order when $(m=2)$ reduced equations with $(l=1)$ are
\begin{eqnarray}
&&n_{12}^{(2)}=C_1|\phi_1^{(1)}|^2,~~~n_{10}^{(2)}=C_6 |\phi_1^{(1)}|^2,\nonumber\\
&&u_{12}^{(2)}=C_2 |\phi_1^{(1)}|^2,~~~~u_{10}^{(2)}=C_7|\phi_1^{(1)}|^2,\nonumber\\
&&n_{22}^{(2)}=C_3|\phi_1^{(1)}|^2,~~~~n_{20}^{(2)}=C_8 |\phi_1^{(1)}|^2,\nonumber\\
&&u_{22}^{(2)}=C_4 |\phi_1^{(1)}|^2,~~~~u_{20}^{(2)}=C_9|\phi_1^{(1)}|^2,\nonumber\\
&&\phi_2^{(2)}=C_5 |\phi_1^{(1)}|^2~~~~~~\phi_0^{(2)}=C_{10} |\phi_1^{(1)}|^2.\label{eq16}
\end{eqnarray}
Finally, the third harmonic modes $(m=3)$ and $(l=1)$ and  with the help of  Eqs. $(\ref{eq13})-(\ref{eq16})$,
give a system of equations, which can be reduced to the following  NLS equation
\begin{eqnarray}
&&~~~i\frac{\partial \Phi}{\partial \tau}+P\frac{\partial^2 \Phi}{\partial \xi^2}+Q|\Phi|^2\Phi=0, \label{eq17}
\end{eqnarray}
where $\Phi=\phi_1^{(1)}$ for simplicity. The dispersion coefficient P
and the nonlinear coefficient Q is given in the appendix.
%%%%%%%%%%%%%%%%%%%%%%%%%%%%%%%%%%%%%%%%%%%%%%%%%%%%%%%%%%%%%%%%%%%%%%%%%%%%%%%%%%%%%%%%%%%%%%%%%%%%%%%%%%%%%%%%%%%%%%%%%%
\section{Modulational instability  and rouge waves}\label{Modulational}
\subsection{MI}
The stability of the DAWs are totally depended on the sign of the product of dispersive term $P$ and nonlinear term $Q$, whereas
the $P$ and $Q$ are function of carrier wave number $k$, negatively charged massive dust temperature (via $\sigma_1$), positively
charged dust temperature (via $\sigma_2$), the ratio of negatively to positively charged dust masses (via $\mu$),
the ratio of positively to negatively  charged dust charge state (via $Z$), the number density of positively charged dust to
negatively charged dust (via $R$), and ion temperature (via $\delta$).
When $PQ>0$, the DAWs are modulationally unstable  against external perturbation (bright envelope solitons exist) and on the other hand
when $PQ<0$, the DAWs are modulationally stable (dark envelope solitons exist) \cite{Kourakis2005,Sukla2002,Schamel2002,Fedele2002}. simultaneously
when $PQ>0$ and ${k_{MI}}<k_c$, the growth rate ($\Gamma_g$) of  MI is
given \cite{Shalini2015}  by
\begin{eqnarray}
&&\Gamma_g=|P|~{k^2_{MI}}\sqrt{\frac{k^2_{c}}{k^2_{MI}}-1}.\label{eq18}
\end{eqnarray}
Here ${k_{MI}}$ is the perturbation wave number and the critical value of the wave number of modulation
$k_c=\sqrt{2Q{|\Phi_o|}^2/P}$, where $\Phi_o$ is the amplitude of the carrier waves.
Hence, the maximum value $\Gamma_{g(max)}$ of $\Gamma_g$ is obtained at ${k_{MI}}=k_c/\sqrt{2}$
and is given by $\Gamma_{g(max)}=|Q||\Phi_0|^2$.
\subsection{rouge wave solution}
The  rogue wave (rational solution) of the NLS Eq. (\ref{eq17}) in the unstable
region ($PQ>0$) can be written \cite{Abdelwahed2016, Ankiewiez2009} as
\begin{eqnarray}
&&\Phi=\sqrt{\frac{2P}{Q}} \left[\frac{4(1+4iP\tau)}{1+16P^2\tau^2+4\xi^2}-1 \right]\mbox{exp}(i2P\tau).\label{eq19}
\end{eqnarray}
The solution (\ref{eq19}) predicts the concentration of the DAWs energy into a small, tiny
region that is caused by the nonlinear behavior of the plasma medium.
\section{Discussion}\label{Discussion}
In our present investigation, from Eq. (\ref{eq14}), we can be observed that there are two possibility of $\omega$,
one for positive sign, and another one for negative sign. For the positive sign, one can recognize it as fast mode ($\omega_f$),
whereas negative sign for slow mode ($\omega_s$). From Figs. $1(a)$ and $1(c)$, both
fast and slow modes can be observed, respectively. From Fig. $1(a)$, fast mode is
exponentially increased with respect to carrier wave number ($k$) but after a particular value of carrier wave number
fast mode remains almost constant. On the other hand slow mode is linearly increased  with the increase of
carrier wave number [see Fig. $1(c)$]. The group velocity ($v_g$) of the DAWs in our plasma system is also function of $\omega$. So the variation of $v_g$
with $k$ and $\mu$ for the fast and slow modes  are not same fashion, which can be observed from Figs. $1(b)$ and $1(d)$, respectively.

The dependency of stability conditions of the  DAWs on  various plasma parameters can be investigated by depicting the ratio of $P/Q$ versus carrier wave number
$k$ for different plasma parameters. When  $P/Q<0$, DAWs are modulationally stable against external perturbation, while  $P/Q>0$, DAWs will be modulationally  unstable
against external perturbations.  When $P/Q\rightarrow\pm\infty$, the corresponding value of $k(=k_c)$
is called critical or threshold wave number for the onset of MI. Figures $2(a)$  and $2(c)$  shows the
variations of $P/Q$ with $k$ and $\mu$ for the fast and slow modes, respectively.
From both figures, it may be manifested that possible stable and unstable region are occurred for DAWs, where
unstable region allows MI growth rate. To highlight the effects of the nonthermal parameter $\kappa$  on the growth rate of MI, Fig. $2(b)$ is
depicted. The growth rate of the MI is so much sensitive to change the
values of $\kappa$, an increase of $\kappa$ value the maximum growth rate is increased. But opposite fashion is observed in Fig. $2(d)$,
which is depicted for different values of ion to electron temperature (via $\delta$). With the increasing
of $\delta$ value the maximum growth rate is decreased [see  Fig. $2(d)$]. So the nonthermal parameter and  ion to electron temperature play
an opposite role on the MI growth rate for our  plasma system.

DARWs are extremely depended on nontharmality parameter ($\kappa$). From Figs. $3(a)-3(c)$, it can be seen that an increase of $\kappa$
value leads to enhancement of DARWs amplitude. That means  large values $\kappa$ stimulated the nonlinearity of the plasma system,
which makes the DARWs taller by  concentrating a significant amount of energy into tiny region. Thus the presence of kappa distributed electrons
changes the structure of the DARWs significantly. But ion to electron temperature (via $\delta$) plays an opposite role to change
the structure of the DARWs [see  Figs. $3(d)-3(f)$]. In the case DARWs amplitude slowly decreases with the increasing of $\delta$.
From this pictures, it can be deduced that energetic ions used to suppressed the nonlinearity of the plasma system which makes the DARWS smaller.

To highlight the effects of ion to negatively charged dust particles number density (via $\mu_i$) in DARWs, Figs. $4(a)-4(c)$ are depicted.
In the case DARWs amplitude get maximized with the increasing of ion number density. Excess number of ion escalating the nonlinearity of plasma
system which manifests gigantic DARWs. Opposite fashion is also observed for positively charged dust charge state to negatively
charged massive dust charge state ratio (via $Z$)  from Figs  $4(d)-4(e)$. In the case, highly positively charged dust are caused to
minimize the nonlinearity of plasma system which leads to decrease the amplitude of DARWs.
So the shape of DARWs is so much sensitive to any change in the ratio of both dust particles charge state.
\section{Conclusion}\label{Conclusion}
We have investigated the amplitude modulation of DAWs packet in an unmagnetized four component
plasma consisting of inertial warm  positively charged and  heavy negatively charged massive dust, as  well
as nonthermal kappa distributed electrons, and isothermal ions. By employing the reductive perturbation method,
a NLS equation is derived, which governs the MI of DAWs, and formation of rogue waves  in the unstable regimes.
To conclude, the findings (rogue waves) of our present investigation should be relevant in space plasma (viz. mesosphere and ionospehre) where kappa
distributed electrons, isothermal ions, as well as positively and negatively charged massive particles are coexist.

\section{Acknowledgment}
N. A. Chowdhury is grateful to the Bangladesh Ministry of Science
and Technology for awarding the National Science and Technology
(NST) Fellowship.
\section{Appendix: Expression of the  coefficients}
\begin{eqnarray}
&&\hspace*{-2.5cm}C_1=\frac{2C_5 k^2 S^2  -(3 \omega^2 k^4+\lambda k^6)}{2S^3},~~~~~C_2=\frac{\omega C_1 S^2 -\omega k^4}{k S^2}, \nonumber\\
&&\hspace*{-2.5cm}C_3=\frac{3\alpha^2\omega^2 k^4 +\theta \alpha^2 k^6+2\alpha C_5 A^2 k^2}{2A^3},~~~~~C_4=\frac{ \omega C_3 A^2-\omega \alpha^2 k^4 }{k A^2},  \nonumber\\
&&\hspace*{-2.5cm}C_5=\frac{A^3 (3\omega^2k^4+\lambda k^6)-2\gamma_2 A^3 S^3+\beta S^3(3\alpha^2 \omega^2 k^4+\theta\alpha^2 k^4)}{2S^2k^2 A^3+2A^3 S^3 (4k^2+\gamma_1)-2\alpha\beta A^2 k^2S^3}, \nonumber\\
&&\hspace*{-2.5cm}C_6=\frac{2 v_g \omega k^3+\lambda k^4+k^2\omega^2-C_{10}S^2}{S^2(v^2_g-\lambda)},~~~~~C_7=\frac{v_g C_6 S^2-2\omega k^3}{S^2}, \nonumber\\
&&\hspace*{-2.5cm}C_8=\frac{2 v_g \omega\alpha^2 k^3+\theta \alpha^2k^4+\alpha^2 k^2 \omega ^2+\alpha C_{10}A^2}{A^2(v^2_g - \theta )},~~~~~C_9=\frac{v_g C_8 A^2 -2\omega\alpha^2 k^3}{A^2}, \nonumber\\
&&\hspace*{-2.5cm}C_{10} =\frac{A^2( 2 v_g \omega k^3+\lambda k^4+k^2 \omega ^2 )(  v^2_g - \theta )+2\gamma_2 A^2 S^2(v^2_g -\theta )( v^2_g -\lambda)-\beta S^2( 2 v_g \omega \alpha^2 k^3+\theta \alpha^2 k^4+\alpha^2 k^2 \omega ^2)(v_g^2 -\lambda)}{\alpha \beta A^2 S^2 (v^2_g -\lambda)+A^2 S^2 (v^2_g - \theta)- \gamma_1 A^2 S^2(v^2_g - \theta)(v^2_g -\lambda)}, \nonumber\
\end{eqnarray}

\begin{eqnarray}
&&\hspace*{-2.5cm}P =\frac{1}{2 AS\omega k^2(A^2+\alpha \beta S^2)}\left[A^3\{(v_g \omega -\lambda k)(\lambda k^3-2\omega v_g k^2+k \omega^2-kS)+(v_g k-\omega)
                                              (\lambda \omega k^2 - 2v_g k \omega^2+\omega^3-k v_g S)\} \right.\nonumber\\
&&\hspace*{-2.5cm}\left.~~~~~~~~~~~~~~-\alpha \beta S^3\{(v_g \omega -\theta k)(\theta k^3-2\omega v_g k^2+k\omega^2+kA)+(v_g k-\omega)(\theta\omega k^2 - 2v_g k\omega^2+\omega^3+kv_gA)\}-A^3S^3\right],\nonumber\\
&& \hspace*{-2.5cm}Q=\frac{A^2S^2}{2\omega k^2(A^2+\alpha \beta S^2)}\left[2\gamma_2(C_5+C_{10})+3\gamma_3-\frac{2\omega k^3(C_2+C_{7})}{S^2}-\frac{2\alpha \beta \omega k^3(C_4+C_9)}{A^2} \right.\nonumber\\
&&\hspace*{-2.5cm}\left.~~~~~~~~~~~~~~~~~~~-\frac{(\omega^2k^2+\lambda k^4)(C_1+C_6)}{S^2} -\frac{(\alpha \beta k^2\omega^2+\alpha \beta \theta k^4)(C_3+C_8)}{A^2}\right].\nonumber\
\end{eqnarray}

\bibliography{mybibfile}

\end{document}